\documentclass[twocolumn, twocolappendix]{aastex631}

\shorttitle{Large-Scale Mapping Observations of DCN and DCO$^{+}$ toward Orion KL}
\shortauthors{Taniguchi et al.}
\graphicspath{{./}{figures/}}

\begin{document}

\title{Large-Scale Mapping Observations of DCN and DCO$^{+}$ toward Orion KL}

\correspondingauthor{Kotomi Taniguchi}
\email{kotomi.taniguchi@nao.ac.jp}

\author[0000-0003-4402-6475]{Kotomi Taniguchi}
\affiliation{National Astronomical Observatory of Japan, National Institutes of Natural Sciences, 2-21-1 Osawa, Mitaka, Tokyo 181-8588, Japan}

\author[0009-0001-6483-7366]{Prathap Rayalacheruvu}
\affiliation{School of Earth and Planetary Sciences, National Institute of Science Education and Research, Jatni 752050, Odisha, India}
\affiliation{Homi Bhabha National Institute, Training School Complex, Anushaktinagar, Mumbai 400094, India}

\author[0000-0003-2386-7427]{Teppei Yonetsu}
\affiliation{Department of Physics, Graduate School of Science, Osaka Metropolitan University, 1-1 Gakuen-cho, Naka-ku, Sakai, Osaka 599-8531, Japan}

\author[0000-0002-4124-797X]{Tatsuya Takekoshi}
\affiliation{Kitami Institute of Technology, 165, Koen-cho, Kitami, Hokkaido 090-8507, Japan} 

\author[0000-0001-6469-8725]{Bunyo Hatsukade}
\affiliation{National Astronomical Observatory of Japan, National Institutes of Natural Sciences, 2-21-1 Osawa, Mitaka, Tokyo 181-8588, Japan}
\affiliation{Graduate Institute for Advanced Studies, SOKENDAI, Osawa, Mitaka, Tokyo 181-8588, Japan}

\author[0000-0002-4052-2394]{Kotaro Kohno}
\affiliation{Institute of Astronomy, Graduate School of Science, The University of Tokyo, 2-21-1 Osawa, Mitaka, Tokyo 181-0015, Japan}
\affiliation{Research Center for the Early Universe, Graduate School of Science, The University of Tokyo, 7-3-1 Hongo, Bunkyo-ku, Tokyo 113-0033, Japan}

\author[0009-0005-5915-1035]{Tai Oshima}
\affiliation{National Astronomical Observatory of Japan, National Institutes of Natural Sciences, 2-21-1 Osawa, Mitaka, Tokyo 181-8588, Japan}
\affiliation{Graduate Institute for Advanced Studies, SOKENDAI, Osawa, Mitaka, Tokyo 181-8588, Japan}

\author[0000-0003-4807-8117]{Yoichi Tamura}
\affiliation{Department of Physics, Graduate School of Science, Nagoya University, Furocho, Chikusa-ku, Nagoya, Aichi 464-8602, Japan}

\author[0000-0002-1413-1963]{Yuki Yoshimura}
\affiliation{Institute of Astronomy, Graduate School of Science, The University of Tokyo, 2-21-1 Osawa, Mitaka, Tokyo 181-0015, Japan}

\author[0009-0003-9025-6121]{V\'{i}ctor G\'{o}mez-Rivera}
\affiliation{Corporaci\'{o}n Mexicana de Investigaci\'{o}n en Materiales S.A. de C.V., M\'{e}xico}
\affiliation{Instituto Nacional de Astrof\'{i}sica, \'{O}ptica y Electr\'{o}nica, Luis Enrique Erro 1, Tonantzintla C.P. 72840, Puebla, M\'{e}xico}

\author[0000-0003-1054-4637]{Sergio Rojas-Garc\'{i}a}
\affiliation{Instituto Nacional de Astrof\'{i}sica, \'{O}ptica y Electr\'{o}nica, Luis Enrique Erro 1, Tonantzintla C.P. 72840, Puebla, M\'{e}xico}

\author[0000-0001-9395-1670]{Arturo I. G\'{o}mez-Ruiz}
\affiliation{Instituto Nacional de Astrof\'{i}sica, \'{O}ptica y Electr\'{o}nica, Luis Enrique Erro 1, Tonantzintla C.P. 72840, Puebla, M\'{e}xico}
\affiliation{Consejo Nacional de Ciencia y Tecnolog\'{i}a, Av. Insurgentes Sur 1582, Col. Cr\'{e}dito Constructor, Demarcaci\'{o}n Territorial Benito Ju\'{a}rez C.P. 03940, Ciudad de M\'{e}xico, M\'{e}xico}

\author{David H. Hughes}
\affiliation{Instituto Nacional de Astrof\'{i}sica, \'{O}ptica y Electr\'{o}nica, Luis Enrique Erro 1, Tonantzintla C.P. 72840, Puebla, M\'{e}xico}

\author{F. Peter Schloerb}
\affiliation{Department of Astronomy, University of Massachusetts, Amherst, MA 01003, USA}

\author[0000-0001-7031-8039]{Liton Majumdar}
\affiliation{School of Earth and Planetary Sciences, National Institute of Science Education and Research, Jatni 752050, Odisha, India}
\affiliation{Homi Bhabha National Institute, Training School Complex, Anushaktinagar, Mumbai 400094, India}

\author[0000-0003-0769-8627]{Masao Saito}
\affiliation{National Astronomical Observatory of Japan, National Institutes of Natural Sciences, 2-21-1 Osawa, Mitaka, Tokyo 181-8588, Japan}
\affiliation{Graduate Institute for Advanced Studies, SOKENDAI, Osawa, Mitaka, Tokyo 181-8588, Japan}

\author[0000-0002-8049-7525]{Ryohei Kawabe}
\affiliation{National Astronomical Observatory of Japan, National Institutes of Natural Sciences, 2-21-1 Osawa, Mitaka, Tokyo 181-8588, Japan}

\begin{abstract}
We present emission maps ($1.5\arcmin\times1.5\arcmin$ scale, corresponding to 0.18 pc) of the DCN ($J=2-1$) and DCO$^+$ ($J=2-1$) lines in the 2 mm band toward the Orion KL region obtained with the 2 mm receiver system named B4R installed on the Large Millimeter Telescope (LMT).
The DCN emission shows a peak at the Orion KL hot core position, whereas no DCO$^+$ emission has been detected there.
The DCO$^+$ emission shows enhancement at the west side of the hot core, which is well shielded from the UV radiation from OB massive stars in the Trapezium cluster.
We have derived the abundance ratio of DCN/DCO$^+$ at three representative positions where both species have been detected.
The gas components with $V_{\rm {LSR}} \approx 7.5-8.7$ km\,s$^{-1}$ are associated with low abundance ratios of $\sim4-6$, whereas much higher abundance ratios ($\sim22-30$) are derived for the gas components with $V_{\rm {LSR}} \approx 9.2-11.6$ km\,s$^{-1}$.
We have compared the observed abundance ratio to our chemical models and found that 
the observed differences in the DCN/DCO$^+$ abundance ratios are explained by different densities.
\end{abstract}


\section{Introduction} \label{sec:intro}

Deuterium fractionation, the D/H ratios in molecules, has been known as a good evolutionary indicator of prestellar and star-forming cores \citep[for reviews;][]{2012AARv..20...56C,2014prpl.conf..859C}.
The cosmic elemental D/H ratio is approximately $1.5\times10^{-5}$ \citep{2003SSRv..106...49L,2003ApJ...587..235O}.
On the other hand, the D/H ratios in molecular species show much higher values than the cosmic elemental ratio, especially in cold ($T\leq10$ K) and dense ($n_{\rm {H}} \geq 10^{6}$ cm$^{-3}$) prestellar cores \citep[e.g.,][]{2005ApJ...619..379C,2006ApJ...645.1198V,2019AA...629A..15R}.

Deuterium fractionation is well known to start with an enhancement of the D/H ratios in several ions such as H$_{3}^{+}$, CH$_{3}^{+}$, and C$_{2}$H$_{2}^{+}$ by the following exothermic isotope-exchange reactions with HD;
\begin{equation} \label{eq:H3}
{\rm {H}}_3^+ + {\rm {HD}} \rightleftharpoons {\rm {H}}_2{\rm {D}}^+ + {\rm {H}}_2 + \Delta\,E \, (232\,{\rm K}),
\end{equation}
\begin{equation} \label{eq:CH3}
{\rm {CH}}_3^+ + {\rm {HD}} \rightleftharpoons {\rm {CH}}_2{\rm {D}}^+ + {\rm {H}}_2 + \Delta\,E \, (390\,{\rm K}),
\end{equation}
and 
\begin{equation} \label{eq:C2H2}
{\rm {C}}_2{\rm {H}}_2^+ + {\rm {HD}} \rightleftharpoons {\rm {C}}_2{\rm {H}}{\rm {D}}^+ + {\rm {H}}_2 + \Delta\,E \,(550\,{\rm K}).
\end{equation}
In cold conditions ($T\leq10$ K), the backward reactions of reactions (\ref{eq:H3}) -- (\ref{eq:C2H2}) are suppressed, and the deuterium is exceeded in these ions.
These ions are precursors to other deuterium species, both molecular and ionic forms.
For example, DCO$^{+}$ is considered to form from H$_{2}$D$^{+}$ in low-temperature regions \citep[$T<30$ K;][]{2014prpl.conf..859C} as follows:
\begin{equation} \label{eq:DCO}
{\rm {H}}_2{\rm {D}}^+ + {\rm {CO}} \rightarrow {\rm {DCO}}^+ + {\rm {H}}_2.
\end{equation}

One main formation route of DCN is considered as the following reaction scheme \citep{1989ApJ...340..906M,2001ApJS..136..579T}:
\begin{equation} \label{eq:DCN1}
{\rm {CH}}_2{\rm {D}}^+ + {\rm {H}}_2 \rightarrow {\rm {CH}}_4{\rm {D}}^+ + h\nu , 
\end{equation}
\begin{equation} \label{eq:DCN2}
{\rm {CH}}_4{\rm {D}}^+ + {\rm {e}}^- \rightarrow {\rm {CHD}} + {\rm {H}}_2 + {\rm {H}},
\end{equation}
followed by 
\begin{equation} \label{eq:DCN3}
{\rm {CHD}} + {\rm {N}} \rightarrow {\rm {DCN}} + {\rm {H}}.
\end{equation}
Another pathway starts with the following reaction:
\begin{equation} \label{eq:DCN4}
{\rm {CH}}_2{\rm {D}}^{+} + {\rm {e}}^- \rightarrow {\rm {CHD}} + {\rm {H}},
\end{equation}
followed by reaction (\ref{eq:DCN3}).
Thus, in cold conditions, DCO$^+$ and DCN are formed via H$_2$D$^+$ and CH$_2$D$^+$, which are produced by reactions (\ref{eq:H3}) and (\ref{eq:CH3}), respectively. 

Several reactions can produce deuterium species in warm regions.
For instance, the following reaction is responsible for the formation of DCO$^{+}$ in warm regions \citep[$> 30$ K;][]{1985ApJ...294L..63A}:
\begin{equation} \label{eq:DCO2}
{\rm {HCO}}^+ + {\rm {D}} \rightarrow {\rm {DCO}}^+ + {\rm {H}}.
\end{equation}

The deuterium fractionation mechanisms that proceed in cold and warm regions have been found to work in active star-forming regions.
\citet{2009AA...508..737P} revealed the deuterium fractionation starting from CH$_2$D$^+$ in warm regions by observations using the APEX and IRAM 30m telescopes toward two clumps in Orion Bar, which is one of the most famous photon-dominated regions (PDRs).
They detected DCN, DCO$^+$, and HDCO, and found significant deuterium fractionation for HCN and H$_2$CO but a low fractionation in HCO$^+$.
They compared the observational results with their chemical simulations and concluded that the deuterium fractionation starting from CH$_2$D$^+$ is important in this region.

\citet{2015AA...579A..80G} investigated chemical evolution using several deuterium species toward 59 sources in high-mass star-forming regions using the Arizona Radio Observatory Submillimeter Telescope (SMT).
They found that the DCN/HCN abundance ratio shows peaks at the hot molecular core stage, whereas the D/H ratios in HNC, HCO$^{+}$, and N$_2$H$^+$ decrease with time evolution.
These results suggest that the deuterium fractionation of DCN can proceed efficiently even in warm or hot regions.
\citet{2022ApJ...925..144S} observed several deuterium species using the Atacama Large Millimeter/submillimeter Array (ALMA) toward the infrared dark cloud (IRDC) G\,14.492-0.139 and found that the emission of DCO$^{+}$ and DCN is signs of star formation activity in the IRDC.
They also found that the observed DCO$^+$/N$_2$D$^+$ abundance ratios in this IRDC clump are lower than those in starless cores in low-mass star-forming regions, implying that the dense cores in this IRDC are warmer and denser than those in low-mass star-forming regions.

\citet {2021ApJ...922..152T} presented observational results of DCO$^+$ and DCN toward two low-mass protostars in the Ophiuchus star-forming region obtained with the ALMA.
They have revealed that different deuterium fractionation mechanisms work around each protostar; deuterium fractionation proceeds in the warm regions ($>30$ K) around Oph-emb9, whereas pathways that can proceed in cold regions ($<30$ K) are efficient around Oph-emb5. 
\citet{2021ApJ...922..152T} found that the UV radiation field may be responsible for the enhancement of these species around Oph-emb9.

The previous studies show that DCO$^+$ and DCN are likely key species to investigate present and past physical conditions in active star-forming regions because they are detected ubiquitously and behave differently in star-forming regions.
Studies to investigate deuterium fractionation have mainly focused on small scales, $i.e.,$ cores and protoplanetary disks \citep[e.g.,][]{2017A&A...606A.125S,2021AJ....161...38O,2021ApJS..257...10C}.
On the other hand, studies of large-scale ($i.e.,$ cloud-scale) deuterium fractionation are scarce so far \citep[e.g., N$_2$H$^+$/N$_2$D$^+$ by][]{2023A&A...675A.190C}.

In this paper, we present the DCN and DCO$^+$ maps toward the Orion KL region \citep[$d=418 \pm 6$ pc;][]{2008PASJ...60..991K}. 
Orion KL is located in the Orion Molecular Cloud-1 (OMC-1). 
Large-scale mapping observations toward the Orion region including OMC-1 were conducted with the Nobeyama 45m radio telescope as a part of a legacy project \citep{2019PASJ...71S...3N}. 
OMC-1 is irradiated by the very strong far-ultraviolet (FUV: 6 eV $< h\nu <$ 13.6 eV) radiation field but has the highest column densities of $^{13}$CO and C$^{18}$O among the Orion A region \citep{2019PASJ...71S...9I}.
Recent studies with the James Webb Space Telescope (JWST) have revealed the 3D structure around this region. 
The Trapezium cluster irradiates the background OMC-1 creating the PDR Orion Bar \citep{2023arXiv231008720P}. 
Although the Trapezium cluster includes some massive OB stars, the effects from the most massive one, the $\theta^1$ Orionis C ($\theta^1$ Ori C), seem to be dominant \citep{2023arXiv231008720P}.
Orion KL is one of the best-studied hot cores rich with molecular lines from complex organic molecules and has been focused by many astrochemical studies \citep[e.g.,][]{1997ApJS..108..301S, 2015A&A...581A..71F, 2015A&A...581A..48G, 2017A&A...605A..76R}.

The data presented in this paper were obtained as a part of commissioning observations of the 2 mm band receiver system, which was named B4R\footnote{\url{http://lmtgtm.org/telescope/instrumentation/instruments/b4r/}}, installed on the Large Millimeter Telescope (LMT) or Gran Telescopio Milim\'{e}trico Alfonso Serrano \citep{2020SPIE11445E..22H}. 
This paper is organized as follows.
In Section \ref{sec:obs}, we explain observational setups and data reduction procedures.
Obtained maps and spectra are presented in Section \ref{sec:res}.
We investigate relationships between the observed DCN/DCO$^+$ abundance ratios and the velocity components in Section \ref{sec:dis1}, and compare the observational results with the chemical simulations in Section \ref{sec:dis2}.
The main conclusions of this paper are summarized in Section \ref{sec:con}.

\section{Observations and Data Reduction} \label{sec:obs}

Observations presented in this paper were obtained during the commissioning of the B4R system from 20 to 29 November 2019.
The data were obtained to make a beam map using the SiO maser line ($J=3-2$, $v=1$).
The coordinate of the map center is ($\alpha_{\rm {J2000}}$, $\delta_{\rm {J2000}}$) = (5$^{\rm {h}}$35$^{\rm {m}}$14\fs16, -5\degr22\arcmin21\farcs50).
The details of the B4R system and results of its commissioning observations are presented in R. Kawabe et al. (2023, in prep.).
The main beam efficiency and beam size at 140 GHz were 59 \% and $10.6\arcsec$, respectively (R. Kawabe et al., 2023).
The frequency setup covers 128.9--131.4 GHz and 142.6--145.1 GHz in LSB and USB, respectively.
The DCO$^{+}$ ($J=2-1$; 144.077289 GHz, $E_{\rm {up}} = 10.4$ K) and DCN ($J=2-1$; 144.8280015 GHz, $E_{\rm {up}} = 10.4$ K)  lines were covered in the USB.
The system noise temperatures ($T_{\rm {sys}}$) were between 100 K and 150 K during the commissioning.

We conducted data reduction with the pipeline (b4rpipe)\footnote{\url{https://github.com/b4r-dev/b4rpipe}} which was developed for the B4R observations.
We used the data in the condition that the pointing accuracy was not degraded by wind during the observations to make final maps.
The final maps were made in the fits format through the pipeline.
The map size is $1.5\arcmin\times1.5\arcmin$, corresponding to a 0.18-pc scale.
A frequency resolution is 76.29 kHz, corresponding to $\sim 0.16$ km\,s$^{-1}$ in the velocity resolution at the observed frequency band.
The beam size of the final maps is $11\arcsec$, corresponding to 0.02 pc at the source distance (418 pc).

After making the fits files with the pipeline, we treated data and conducted further data reduction and analyses using the Common Astronomy Software Applications (CASA) package \citep{2022PASP..134k4501C}.
We corrected the position offset.
The 5th-order polynomial fitting calculated from line-free channels was applied for the baseline fitting.
The detailed methods of the procedure to make final maps will be explained in T. Yonetsu et al. (in prep.).

When constructing moment maps in CASA, we applied the sigma-clipping method, because several velocity components are mixed in this region.
We made masks for emissions above $3\sigma$ for the 3D cube fits using the {\it {immath}} task, and smoothed the masks with a single-beam scale (11\arcsec) using the {\it {imsmooth}} task.
The $3\sigma$ noise thresholds are 0.27 K and 0.25 K for the DCN and DCO$^+$ maps, respectively.
We applied the generated masks for the original 3D cube data using the {\it {immath}} and obtained the 3D cube data.
Using the new 3D cube data applied to the sigma-clipping method, we made moment maps presented in Section \ref{sec:map}.

\section{Results} \label{sec:res}

\subsection{Maps of DCN and DCO$^+$} \label{sec:map}

\begin{figure*}
 \begin{center}
  \includegraphics[bb = 0 15 400 650, scale = 0.8]{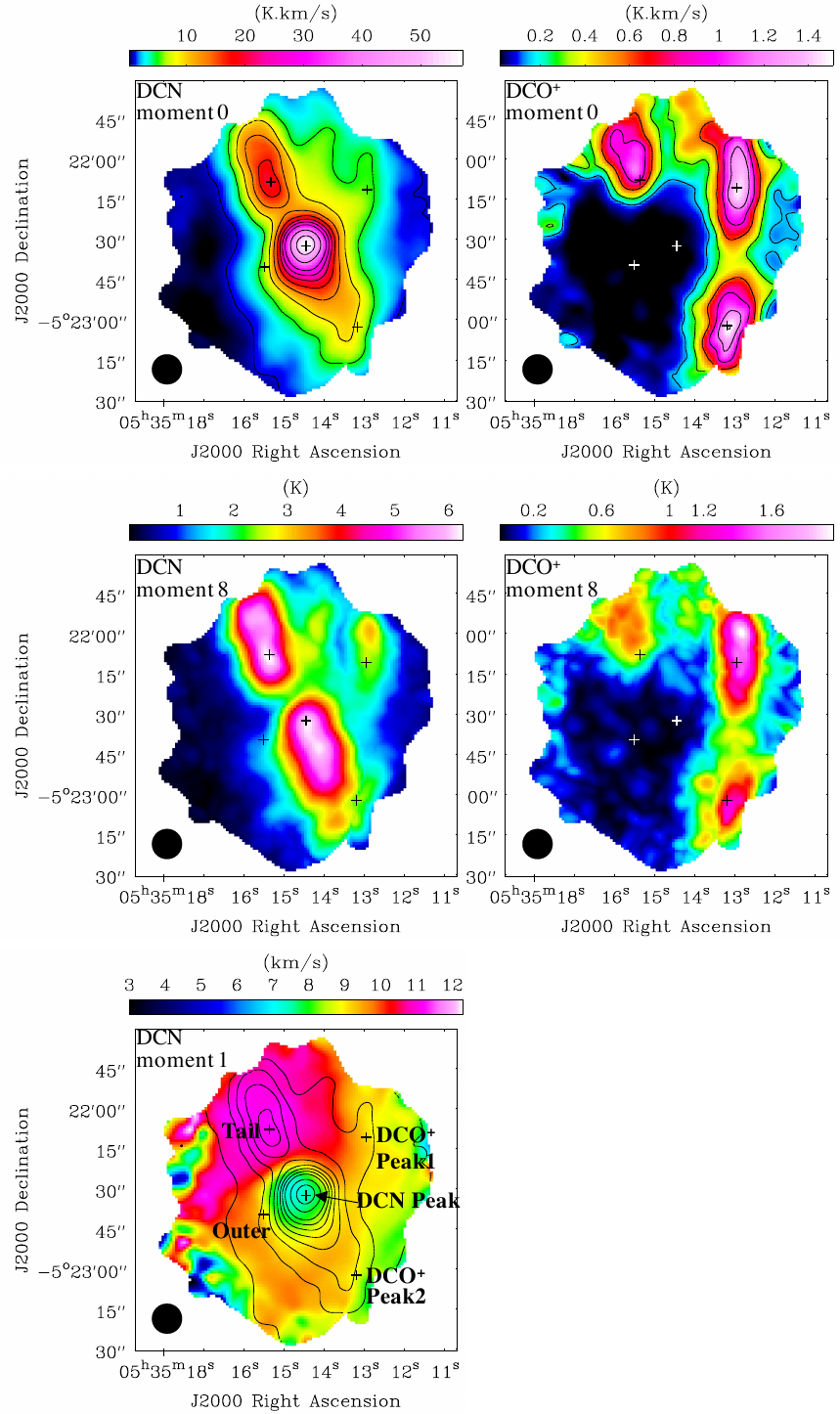}
 \end{center}
\caption{Integrated-intensity of DCN and DCO$^{+}$ (moment 0 maps; top panels), peak intensity maps of DCN and DCO$^{+}$ (moment 8 maps; middle panels), and velocity map of DCN (moment 1 map; bottom panel) toward the Orion KL region. The black contours in the DCN moment 1 map show the same one as shown in its moment 0 map. The black crosses indicate five representative positions where we analyze spectra. The names of each position are indicated in the DCN moment 1 map. \label{fig:mom}}
\end{figure*}

Figure \ref{fig:mom} shows integrated-intensity maps (moment 0 maps; top panels) and peak intensity maps (moment 8 maps; middle panels) of the DCN and DCO$^+$ lines, and a velocity map of DCN (moment 1 map; bottom panel).
Based on the moment 0 maps, we identified five representative positions; DCN Peak, DCO$^{+}$ Peak1, DCO$^{+}$ Peak2, Tail, and Outer.
The coordinate, H$_2$ column density, indicated as $N$(H$_2$), dust temperature, and visual extinction ($A_{\rm {V}}$) at each position are summarized in Table \ref{tab:phy}.
The values of $N$(H$_2$) and dust temperature are obtained from maps of \citet{2021AA...651A..36S}.
The H$_2$ column density map has an angular resolution of 8\arcsec.
This map was made using SPIRE 160, 250, 350, and 500 $\micron$ data from {\it {Herschel}} Gould Belt Survey \citep[HGBS;][]{2010A&A...518L.102A}, and ArT\'{e}MiS 350 and 450 $\micron$ data.
The fits file of the dust temperature has an angular resolution of SPIRE 250 $\micron$ (18.2\arcsec).
This map was made by combining data of ArT\'{e}MiS 350 and 450 $\micron$, SPIRE 350 and 500 $\micron$, and SPIRE 160 and 250 $\micron$ data from HGBS.

\begin{figure*}
 \begin{center}
  \includegraphics[bb = 0 15 400 170, scale = 0.8]{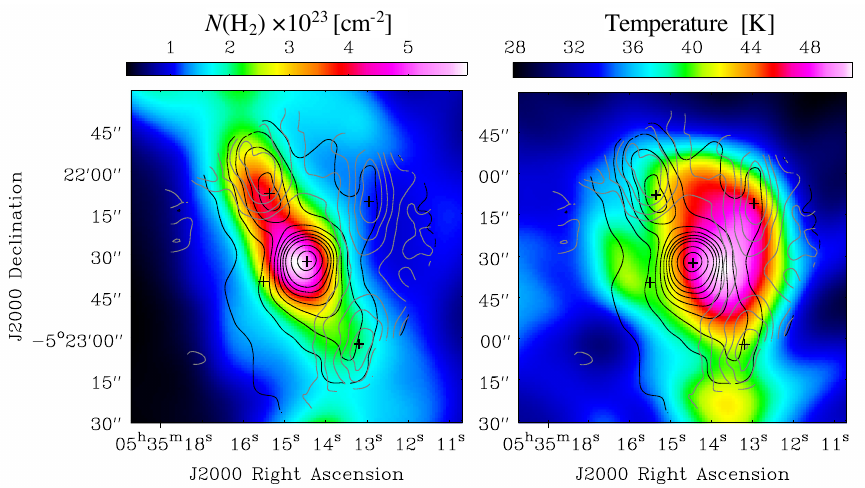}
 \end{center}
\caption{The molecular hydrogen column density (left) and dust temperature (right) maps overlaid by moment 0 maps of DCN (black contours) and DCO$^{+}$ (gray contours). The contours are the same ones as the top panels of Figure \ref{fig:mom}. The black crosses are the same ones in Figure \ref{fig:mom}. \label{fig:phy}}
\end{figure*}

We calculated the visual extinction ($A_{\rm {v}}$) from the H$_{2}$ column density using the following formulae \citep{2021AA...645A..27G}: 
\begin{equation}
A_{\rm {V}} = \frac{2N({\rm {H}}_2)}{1.8\times10^{21}} \,({\rm {mag}}).
\end{equation}
Here, we assumed that all of the hydrogen nuclei are fixed in the molecular form (H$_2$).
These five positions are indicated in the maps as black crosses.
Figure \ref{fig:phy} shows maps of $N$(H$_{2}$) and dust temperature obtained from \citet{2021AA...651A..36S}, overlaid by moment 0 maps of DCN (black contours) and DCO$^{+}$ (gray contours).

Moment 0 maps of the two deuterium species clearly show different spatial distributions.
The DCN emission shows a peak associated with the hot core position and extends from northeast to southwest, similar to the $N$(H$_2$) map (left panel of Figure \ref{fig:phy}). 
On the other hand, the DCO$^+$ emission is mainly enhanced at the western parts from the hot core position.
In addition, the DCO$^+$ emission is associated with Tail as well as DCN.
Several recombination lines from some atoms ($e.g.,$ hydrogen and carbon), which trace \ion{H}{2} regions, have been detected at the east side from the Orion KL hot core position \citep[e.g.,][]{2021A&A...647L...7G}.
Thus, the DCO$^{+}$ emission is enhanced in the shield regions against the UV radiation from the \ion{H}{2} region.
Enhancement in the temperature at the west side near DCO$^+$ Peak1 is likely heated by the Orion KL hot core source, rather than the UV radiation from the \ion{H}{2} region.

\begin{deluxetable*}{lccccc}
\tablecaption{The H$_{2}$ column density and temperature at the five representative positions \label{tab:phy}}
\tablewidth{0pt}
\tablehead{
\colhead{Position} & \colhead{R.A. (J2000)} & \colhead{Decl. (J2000)} & \colhead{$N$(H$_2$)} & \colhead{$T$} & \colhead{$A_{\rm {v}}$} \\
\colhead{} & \colhead{} & \colhead{} & \colhead{($\times10^{23}$ cm$^{-2}$)} & \colhead{(K)} & \colhead{(mag)}
}
\startdata
DCN Peak &5$^{\rm {h}}$35$^{\rm {m}}$14\fs449 & -5\degr22\arcmin32\farcs531 & 5.37 (0.50) & 46.98 (1.59) & 597 (55) \\
DCO$^{+}$ Peak1 & 5$^{\rm {h}}$35$^{\rm {m}}$12\fs922 & -5\degr22\arcmin11\farcs515 & 1.12 (0.09) & 45.30 (1.85) & 124 (10) \\
DCO$^{+}$ Peak2 & 5$^{\rm {h}}$35$^{\rm {m}}$13\fs161 & -5\degr23\arcmin02\farcs899 & 1.98 (0.13) & 40.41 (1.06) & 220 (14) \\
Tail & 5$^{\rm {h}}$35$^{\rm {m}}$15\fs328 & -5\degr22\arcmin08\farcs573 & 3.71 (0.34) & 40.78 (1.26) & 413 (38) \\
Outer & 5$^{\rm {h}}$35$^{\rm {m}}$15\fs476 & -5\degr22\arcmin40\farcs412 & 2.67 (0.65) & 40.12 (0.28) & 296 (72) \\
\enddata
\tablecomments{The H$_{2}$ column density and temperature are obtained from fits files obtained from \citet{2021AA...651A..36S}. The figures in parentheses refer to the standard deviation obtained from the fits files.} \end{deluxetable*}

\subsection{Spectral Analyses} \label{sec:spectra}

\begin{figure*}
 \begin{center}
  \includegraphics[bb = 0 25 400 715, scale = 0.95]{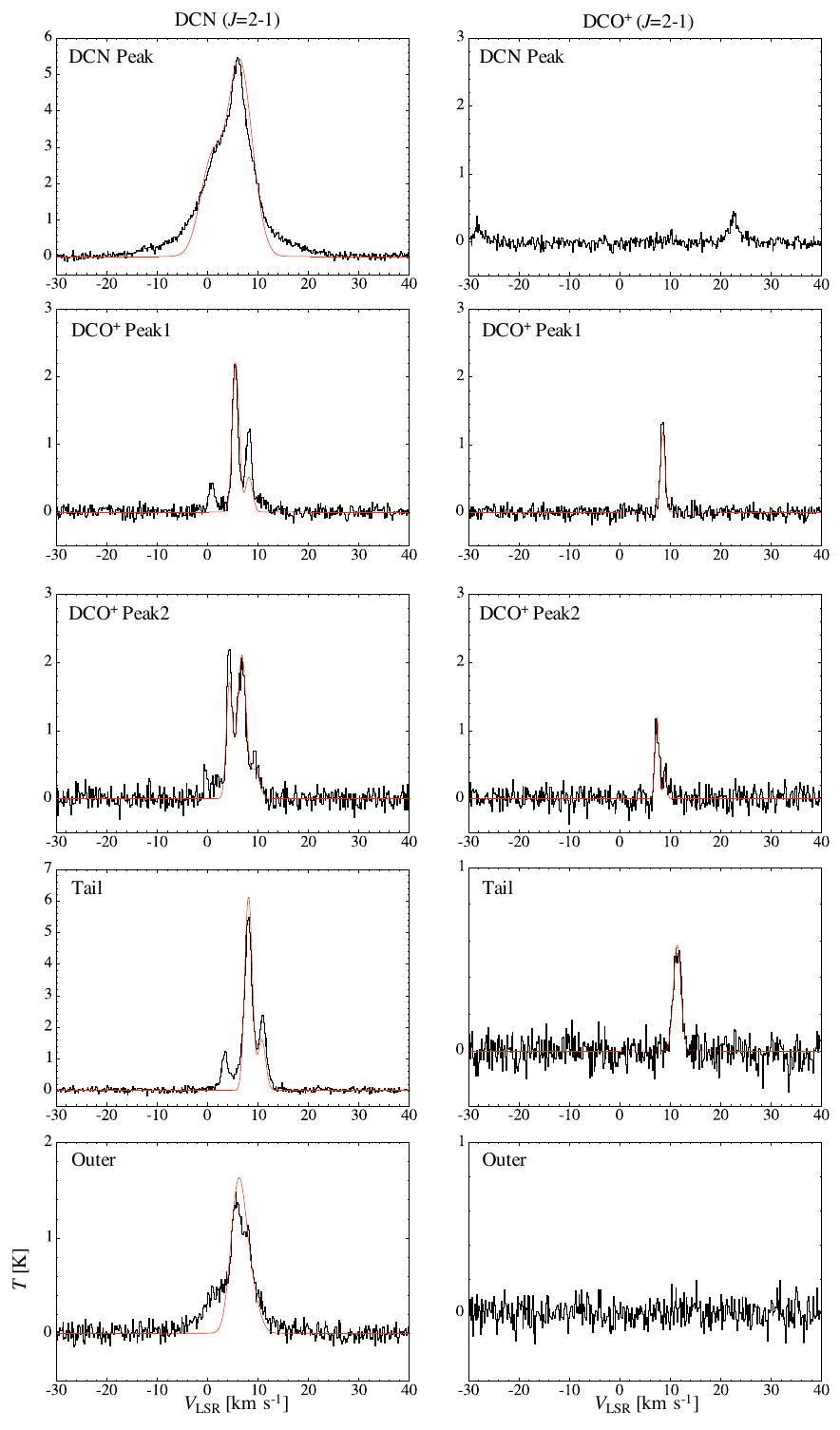}
 \end{center}
\caption{Spectra of the DCN $J=2-1$ line (left panels) and DCO$^{+}$ $J=2-1$ line (right panels) at the five positions. Black and red curves are observed spectra and the best fitting results with an assumed excitation temperature of 40 K, respectively. \label{fig:spec}}
\end{figure*}

Black curves in Figure \ref{fig:spec} show the observed spectra of the DCN (left) and DCO$^{+}$ (right) lines at the five representative positions.
These spectra are extracted over the beam size ($11\arcsec$), 
and the center coordinates are set at the values summarized in Table \ref{tab:phy}.
The DCN ($J=2-1$) line has been detected from all of the positions, whereas the DCO$^{+}$ ($J=2-1$) line has been detected from the three positions except for DCN Peak and Outer, both of which are illuminated by the UV radiation from the massive star(s).

We analyzed spectra with the Markov Chain Monte Carlo (MCMC) method in the CASSIS software \citep{2015sf2a.conf..313V}.
In this method, column density ($N$), excitation temperature ($T_{\rm {ex}}$), velocity component ($V_{\rm {LSR}}$), line width (FWHM) are treated as free parameters.
We assumed the thermodynamic equilibrium (LTE) condition.
This assumption is reasonable here because lines of deuterium species are usually optically thin.
We have only one transition line for each species, and we could not determine their column density and excitation temperature simultaneously.
We then analyzed spectra changing excitation temperatures from 10 K to 50 K with the 5-K step.
The lower limit of the temperature was determined because the deuterium fractionation prefers cold regions (Section \ref{sec:intro}), and its upper limit was determined from the dust temperature map (Figure \ref{fig:phy} and Table \ref{tab:phy}).
Regarding DCN Peak, we changed the excitation temperature from 40 K to 50 K, judging from the dust temperature map (right panel of Figure \ref{fig:phy}) and broad line widths.
The red curves in Figure \ref{fig:spec} show the best-fitting model with an assumed excitation temperature of 40 K.
Even if we changed the assumed excitation temperatures from 10 K to 50 K, the best-fitting models show similar spectral features and there is no significant difference.
We applied two-velocity-component fitting for the spectra at DCN Peak and DCO$^{+}$ Peak2.
At DCO$^+$ Peak1, DCO$^+$ Peak2, and Tail, we analyzed the DCO$^+$ spectra at first and utilized the obtained velocity components to constrain the velocity ranges of the DCN line at each position for secure line identification.
The hyperfine splitting due to nitrogen nuclei and deuterium is taken into account.

The results obtained by the MCMC method are summarized in Table \ref{tab:mcmc}.
The velocity components are consistent between the DCN and DCO$^+$ lines at each position, suggesting that they trace similar regions.
The DCN/DCO$^+$ abundance ratios at each position are summarized in the last column of Table \ref{tab:mcmc}.
Table \ref{tab:abund} summarizes fractional abundances denoted as $X$(molecules), defined by $N$(molecules)/$N$(H$_2$), at each position.
The optical depths of the DCN line are less than 0.73, and those of the DCO$^+$ line are less than 0.29, in the assumed excitation temperature range and at all of the positions.
Thus, these lines are optically thin in all of the cases.

The wing emission in the DCN spectra is likely caused by turbulence or outflows.
We did not include these effects in the spectral fitting procedure.
Another velocity component around $V_{\rm {LSR}}\approx 1-4$ km\,s$^{-1}$ is seen in the DCN spectra at DCO$^+$ Peak1 and Tail. 
We tried fitting these lines as another velocity component, but the lines could not be fitted simultaneously. 
Thus, these velocity components are contamination of another molecular line or emission from completely different regions.
The following discussions and our conclusions are not affected by these components, because we will discuss mainly using the DCN/DCO$^+$ abundance ratios of the velocity components which are common in both species.

\begin{deluxetable*}{lcccccccc}
\tabletypesize{\scriptsize}
\tablecaption{Results of the MCMC analysis \label{tab:mcmc}}
\tablewidth{0pt}
\tablehead{
\colhead{} & \multicolumn{3}{c}{DCN} & \colhead{} & \multicolumn{3}{c}{DCO$^+$} & \colhead{} \\
\cline{2-4} \cline{6-8}
\colhead{Position} & \colhead{$N$ ($\times10^{12}$ cm$^{-2}$)} & \colhead{FWHM (km\,s$^{-1}$)} & \colhead{$V_{\rm {LSR}}$ (km\,s$^{-1}$)} & \colhead{} & \colhead{$N$ ($\times10^{12}$ cm$^{-2}$)} & \colhead{FWHM (km\,s$^{-1}$)} & \colhead{$V_{\rm {LSR}}$ (km\,s$^{-1}$)} & \colhead{DCN/DCO$^+$}
}
\startdata
\multicolumn{9}{l}{$T_{\rm {ex}}=10$ K} \\
DCO$^+$ Peak1 & 15 (2) & 1.38 (1) & 8.52 (1) & & 2.1 (2) & 1.03 (1) & 8.69 (1) & 7 (1) \\		
DCO$^+$ Peak2 & 8.7 (9) & 1.40 (1) & 7.30 (1) & & 1.8 (2) & 0.86 (1) & 7.55 (1) & 4.7 (7) \\ 
			     & 13 (1) & 1.95 (1) & 9.70 (1) & & 0.46 (5) & 0.80 (1) & 9.15 (1) & 28 (4) \\
Tail & \multicolumn{3}{c}{\it {Spectra could not be fitted under the assumption.}} & & 1.5 (2) & 1.70 (2) & 11.56 (1) \\
Outer & 19 (3) & 2.8 (7) & 9.1 (3) & & ... & ... & ... \\
\cline{1-9}
\multicolumn{9}{l}{$T_{\rm {ex}}=15$ K} \\
DCO$^+$ Peak1 & 13 (1) & 1.43 (5) & 8.52 (1) & & 2.0 (2) & 1.04 (1) & 8.69 (1) & 6.3 (9) \\
DCO$^+$ Peak2 & 8.0 (8) & 1.40 (1) & 7.30 (1) & & 1.7 (2) & 0.88 (1) & 7.55 (1) & 4.7 (7) \\
			     & 12 (1) & 1.98 (2) & 9.70 (1) & & 0.44 (5) & 0.80 (1) & 9.15 (1) & 27 (4) \\
Tail 			     & 56 (7) & 1.66 (5) & 11.10 (5) & & 1.5 (2) & 1.72 (2) & 11.55 (1) & 37 (6) \\
Outer 		     & 17 (3) & 2.7 (8) & 9.0 (2) & & ... & ... & ... & \\
\cline{1-9}
\multicolumn{9}{l}{$T_{\rm {ex}}=20$ K} \\
DCO$^+$ Peak1 & 13 (1) & 	1.49 (1) & 8.53 (1) & & 2.1 (2) & 1.05 (1) & 8.69 (1) & 5.9 (8) \\
DCO$^+$ Peak2 & 8.5 (9) & 1.40 (1) & 7.30 (1) & & 1.8 (2) & 0.89 (1) & 7.55 (1) & 4.7 (7) \\
			     & 13 (1) & 2.00 (2) & 9.70 (1) & & 0.48 (5) & 0.80 (1) & 9.15 (1) & 26 (4) \\
Tail 			     & 50 (8) & 1.7 (3) & 11.07 (9) & & 1.7 (2) & 1.50 (1) & 11.56 (1) & 30 (6) \\
Outer 		     & 18 (3) & 3.2 (4) & 9.0 (3) & & ... & ... & ... & \\
\cline{1-9}
\multicolumn{9}{l}{$T_{\rm {ex}}=25$ K} \\
DCO$^+$ Peak1 & 14 (1) & 1.51 (1) & 8.54 (1) & & 2.3 (2) & 1.05 (1) & 8.69 (1) & 5.8 (8) \\
DCO$^+$ Peak2 & 9.3 (9) & 1.40 (1) & 7.29 (1) & & 2.0 (2) & 0.89 (1) & 7.55 (1) & 4.7 (7) \\
			     & 14 (1) & 2.02 (2) & 9.70 (1) & & 0.53 (5) & 0.80 (1) & 9.15 (1) & 26 (4) \\
Tail 			     & 56 (8) & 1.8 (1) & 11.07 (5) & & 1.8 (2) & 1.48 (3) & 11.56 (2) & 30 (5) \\
Outer		     & 20 (3) & 3.2 (4) & 8.9 (2) & & ... & ... & ... & \\
\cline{1-9}
\multicolumn{9}{l}{$T_{\rm {ex}}=30$ K} \\
DCO$^+$ Peak1 & 15 (2) & 1.52 (1) & 8.54 (1) & & 2.6 (3) & 1.05 (1) & 8.69 (1) & 5.8 (8) \\
DCO$^+$ Peak2 & 10 (1) & 1.40 (1) & 7.29 (1) & & 2.2 (2) & 0.89 (1) & 7.55 (1) & 4.7 (7) \\
			    & 15 (2) & 2.01 (2) & 9.70 (1) & & 0.58 (6) & 0.80 (1) & 9.15 (1) & 26 (4) \\
Tail                & 57 (6) & 1.88 (2) & 11.06 (4) & & 2.0 (2) & 1.73 (2) & 11.55 (1) & 28 (4) \\
Outer 		    & 23 (3) & 3.49 (1) & 9.05 (8) & 	& ... & ... & ... & \\		     	
\cline{1-9}
\multicolumn{9}{l}{$T_{\rm {ex}}=35$ K} \\
DCO$^+$ Peak1 & 16 (2) & 1.53 (2) & 8.54 (1) & & 2.8 (3) & 1.05 (1) & 8.69 (1) & 5.7 (8) \\
DCO$^+$ Peak2 & 11 (1) & 1.40 (1) & 7.29 (1) & & 2.4 (2) & 0.89 (1) & 7.55 (1) & 4.6 (7) \\
			     &  17 (2) & 2.02 (1) & 9.70 (1) & & 0.64 (6) & 0.80 (1) & 9.15 (1) & 26 (4) \\
Tail			     & 56 (7) & 1.58 (3) & 11.09 (7) & & 2.2 (2) & 1.73 (1) & 11.55 (1) & 25 (4) \\
Outer 		     & 25 (3) & 3.48 (3) & 9.10 (6) &  & ... & ... & ... & \\		
\cline{1-9}
\multicolumn{9}{l}{$T_{\rm {ex}}=40$ K} \\
DCN Peak & 73 (7) & 5.50 (1) & 3.90 (1) &  & ... & ... & ... \\
		  &  114 (12) & 4.50 (1) & 9.30 (1) & & ... & ... & ... \\
DCO$^+$ Peak1 & 18 (2) & 1.53 (1) & 8.54 (1) & & 3.1 (3) & 1.05 (1) & 8.69 (1) & 5.7 (8) \\
DCO$^+$ Peak2 & 12 (1) & 1.40 (1) & 7.29 (1) & & 2.6 (3) & 0.89 (1) & 7.55 (1) & 4.6 (7) \\
			     & 18 (2) & 2.02 (2) & 9.70 (1) & & 0.70 (7) & 0.80 (1) & 9.15 (1) & 26 (4) \\
Tail 			     & 56 (8) & 1.64 (2) & 11.10 (5) & & 2.4 (2) & 1.73 (1) & 11.55 (1) & 23 (4) \\
Outer 		     & 28 (3) & 3.50 (1) & 9.07 (2) & & ... & ... & ... & \\	
\cline{1-9}
\multicolumn{9}{l}{$T_{\rm {ex}}=45$ K} \\
DCN Peak & 74 (7) & 5.50 (1) & 4.40 (1) &  & ... & ... & ... \\	
		  & 120 (12) & 4.50 (1) & 9.30 (1) &  & ... & ... & ... \\
DCO$^+$ Peak1 & 19 (2) & 1.53 (1) & 8.54 (1) & & 3.3 (3) & 1.06 (1) & 8.69 (1) & 5.7 (8) \\
DCO$^+$ Peak2 & 13 (1) & 1.40 (1) & 7.29 (1) & & 2.9 (3) & 0.89 (1) & 7.55 (1) & 4.6 (7) \\
		  & 20 (2) & 2.02 (1) & 9.70 (1) & & 0.78 (8) & 0.80 (1) & 9.15 (1) & 25 (4) \\
Tail	    & 58 (6) & 1.58 (3) & 11.12 (9) & & 2.6 (3) & 1.73 (1) & 11.55 (1) & 22 (3) \\
Outer 		& 30 (3) & 3.50 (1) & 9.07 (1) &	& ... & ... & ... & \\
\cline{1-9}
\multicolumn{9}{l}{$T_{\rm {ex}}=50$ K} \\
DCN Peak & 84 (8) & 5.50 (1) & 3.90 (1) &  & ... & ... & ... \\	
		  &  130 (13) & 4.50 (1) & 9.35 (1) & & ... & ... & ... \\	   
DCO$^+$ Peak1 & 21 (2) & 1.54 (1) & 8.54 (1) & & 3.6 (4) & 1.06 (1) & 8.69 (1) & 5.7 (8) \\
DCO$^+$ Peak2 & 14 (1) & 1.40 (1) & 7.29 (1) & & 3.1 (3) & 0.89 (1) & 7.55 (1) & 4.6 (7) \\
		   & 21 (2) & 2.02 (2) & 9.70 (1) & & 0.83 (8) & 0.80 (1) & 9.15 (1) & 26 (4) \\
Tail 	      & 67 (8) & 1.58 (2) & 11.09 (5) & & 2.9 (3) & 1.73 (1) & 11.55 (1) & 24 (4) \\
Outer 		  & 32 (3) & 3.50 (1) & 9.07 (2) &	& ... & ... & ... & \\		      		           		               		        		      		  
\enddata
\tablecomments{The figures in parentheses refer to the standard deviation derived from the MCMC analysis, expressed in units of the last significant digits. The errors of the column densities include the absolute calibration error of 10\%.}
\end{deluxetable*}

\begin{deluxetable}{lcc}
\tabletypesize{\scriptsize}
\tablecaption{Fractional abundances of DCN and DCO$^+$ at each position \label{tab:abund}}
\tablewidth{0pt}
\tablehead{
\colhead{Position} & \colhead{$X$(DCN)} & \colhead{$X$(DCO$^+$)}
}
\startdata
\multicolumn{3}{l}{$T_{\rm {ex}}=10$ K} \\
DCO$^{+}$ Peak1 & ($1.3 \pm 0.2$)$\times10^{-10}$ & ($1.9 \pm 0.2$)$\times10^{-11}$ \\
DCO$^{+}$ Peak2 & ($1.1 \pm 0.1$)$\times10^{-10}$ & ($1.2 \pm 0.1$)$\times10^{-11}$ \\
Tail            &   ...                             & ($4.2 \pm 0.6$)$\times10^{-12}$ \\ 
Outer           & ($7 \pm 2$)$\times10^{-11}$   &  ...                              \\
\cline{1-3}
\multicolumn{3}{l}{$T_{\rm {ex}}=15$ K} \\
DCO$^{+}$ Peak1 & ($1.1 \pm 0.2$)$\times10^{-10}$ & ($1.8 \pm 0.2$)$\times10^{-11}$ \\
DCO$^{+}$ Peak2 & ($10 \pm 1$)$\times10^{-11}$ & ($1.1 \pm 0.1$)$\times10^{-11}$ \\
Tail            & ($1.5 \pm 0.2$)$\times10^{-10}$ & ($4.1 \pm 0.6$)$\times10^{-12}$ \\ 
Outer           & ($6 \pm 2$)$\times10^{-11}$   &  ...                               \\
\cline{1-3}
\multicolumn{3}{l}{$T_{\rm {ex}}=20$ K} \\
DCO$^{+}$ Peak1 & ($1.1 \pm 0.1$)$\times10^{-10}$ & ($1.9 \pm 0.2$)$\times10^{-11}$ \\
DCO$^{+}$ Peak2 & ($1.1 \pm 0.1$)$\times10^{-10}$ & ($1.2 \pm 0.1$)$\times10^{-11}$ \\
Tail            & ($1.4 \pm 0.3$)$\times10^{-10}$ & ($4.5 \pm 0.6$)$\times10^{-12}$ \\ 
Outer           & ($7 \pm 2$)$\times10^{-11}$   &  ...                               \\
\cline{1-3}
\multicolumn{3}{l}{$T_{\rm {ex}}=25$ K} \\
DCO$^{+}$ Peak1 & ($1.2 \pm 0.2$)$\times10^{-10}$ & ($2.1 \pm 0.3$)$\times10^{-11}$ \\
DCO$^{+}$ Peak2 & ($1.2 \pm 0.1$)$\times10^{-10}$ & ($1.3 \pm 0.1$)$\times10^{-11}$ \\
Tail            & ($1.5 \pm 0.3$)$\times10^{-10}$ & ($5.0 \pm 0.7$)$\times10^{-12}$ \\ 
Outer           & ($7 \pm 2$)$\times10^{-11}$   &  ...                               \\
\cline{1-3}
\multicolumn{3}{l}{$T_{\rm {ex}}=30$ K} \\
DCO$^{+}$ Peak1 & ($1.3 \pm 0.2$)$\times10^{-10}$ & ($2.3 \pm 0.3$)$\times10^{-11}$ \\
DCO$^{+}$ Peak2 & ($1.3 \pm 0.1$)$\times10^{-10}$ & ($1.4 \pm 0.2$)$\times10^{-11}$ \\
Tail            & ($1.5 \pm 0.2$)$\times10^{-10}$ & ($5.4 \pm 0.7$)$\times10^{-12}$ \\ 
Outer           & ($9 \pm 2$)$\times10^{-11}$   &  ...                               \\
\cline{1-3}
\multicolumn{3}{l}{$T_{\rm {ex}}=35$ K} \\
DCO$^{+}$ Peak1 & ($1.4 \pm 0.2$)$\times10^{-10}$ & ($2.3 \pm 0.3$)$\times10^{-11}$ \\
DCO$^{+}$ Peak2 & ($1.4 \pm 0.1$)$\times10^{-10}$ & ($1.5 \pm 0.2$)$\times10^{-11}$ \\
Tail            & ($1.5 \pm 0.2$)$\times10^{-10}$ & ($6.0 \pm 0.8$)$\times10^{-12}$ \\ 
Outer           & ($9 \pm 3$)$\times10^{-11}$   &  ...                               \\
\cline{1-3}
\multicolumn{3}{l}{$T_{\rm {ex}}=40$ K} \\
DCN Peak        & ($3.5 \pm 0.3$)$\times10^{-10}$ &   ...                             \\
DCO$^{+}$ Peak1 & ($1.6 \pm 0.2$)$\times10^{-10}$ & ($2.8 \pm 0.4$)$\times10^{-11}$ \\
DCO$^{+}$ Peak2 & ($1.5 \pm 0.2$)$\times10^{-10}$ & ($1.7 \pm 0.2$)$\times10^{-11}$ \\
Tail            & ($1.5 \pm 0.3$)$\times10^{-10}$ & ($6.5 \pm 0.9$)$\times10^{-12}$ \\ 
Outer           & ($1.0 \pm 0.3$)$\times10^{-10}$   &  ...                             \\
\cline{1-3}
\multicolumn{3}{l}{$T_{\rm {ex}}=45$ K} \\
DCN Peak        & ($3.6 \pm 0.3$)$\times10^{-10}$ &   ...                             \\
DCO$^{+}$ Peak1 & ($1.7 \pm 0.2$)$\times10^{-10}$ & ($3.0 \pm 0.4$)$\times10^{-11}$ \\
DCO$^{+}$ Peak2 & ($1.7 \pm 0.2$)$\times10^{-10}$ & ($1.8 \pm 0.2$)$\times10^{-11}$ \\
Tail            & ($1.6 \pm 0.2$)$\times10^{-10}$ & ($7.1 \pm 0.9$)$\times10^{-12}$ \\ 
Outer           & ($1.1 \pm 0.3$)$\times10^{-10}$   &  ...                            \\
\cline{1-3}
\multicolumn{3}{l}{$T_{\rm {ex}}=50$ K} \\
DCN Peak        & ($4.0 \pm 0.4$)$\times10^{-10}$ &   ...                             \\
DCO$^{+}$ Peak1 & ($1.8 \pm 0.2$)$\times10^{-10}$ & ($3.2 \pm 0.4$)$\times10^{-11}$ \\
DCO$^{+}$ Peak2 & ($1.8 \pm 0.2$)$\times10^{-10}$ & ($2.0 \pm 0.2$)$\times10^{-11}$ \\
Tail            & ($1.8 \pm 0.3$)$\times10^{-10}$ & ($7.7 \pm 1.0$)$\times10^{-12}$ \\ 
Outer           & ($1.2 \pm 0.3$)$\times10^{-10}$   &  ...                            \\
\enddata
\tablecomments{The figures in parentheses refer to the standard deviation.}
\end{deluxetable}

\section{Discussions} \label{sec:dis}

\subsection{Characteristics of the observed DCN/DCO$^+$ abundance ratios} \label{sec:dis1}

\begin{figure}
 \begin{center}
  \includegraphics[bb = 50 25 150 190, scale = 0.85]{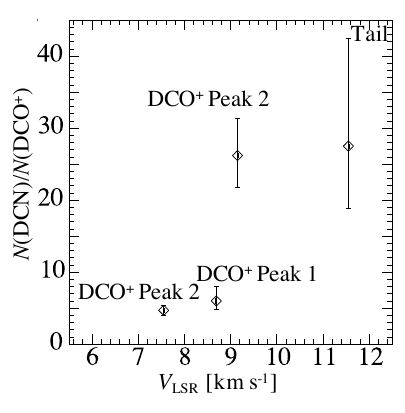}
 \end{center}
\caption{Comparison of the $N$(DCN)/$N$(DCO$^+$) ratio and the velocity component. The velocities are values derived by fitting the DCO$^{+}$ line. \label{fig:ratio_v}}
\end{figure}

We found that the DCN/DCO$^+$ abundance ratios are divided into two groups (Table \ref{tab:mcmc}); the ratio of $\sim4-7$ (low) and $\sim 22-30$ (high).
At DCO$^+$ Peak2, we found both components in the different velocity components, which means that we need to investigate the DCN/DCO$^{+}$ abundance ratios in the three dimensions.

Figure \ref{fig:ratio_v} shows a dependence of the DCN/DCO$^+$ abundance ratio on the velocity component.
In Figure \ref{fig:ratio_v}, we utilized the velocity components derived by fitting the DCO$^+$ line (Table \ref{tab:mcmc}).
The lower abundance ratios are associated with the gas of $V_{\rm {LSR}} \approx 7.5-8.7$ km\,s$^{-1}$, whereas the higher abundance ratios are accompanied with the velocity components of $\approx 9.2-11.6$ km\,s$^{-1}$.
These results imply that gases with different physical conditions are mixed. 

\citet{2019Natur.565..618P} found that adiabatic expansion of the hot gas has swept up the surrounding gas, and the velocity of the expanding shell has a velocity of 13 km\,s$^{-1}$.
This expanding shell is formed by the most massive star in the Trapezium cluster ($\theta^1$ Ori C).
The massive star is located at ($\alpha_{\rm {J2000}}$, $\delta_{\rm {J2000}}$) = (5$^{\rm {h}}$35$^{\rm {m}}$16\fs46, -5\degr23\arcmin22\farcs8).
The velocity of the expanding shell is close to the gas components with the higher DCN/DCO$^+$ abundance ratios ($V_{\rm {LSR}} \approx 9.2-11.6$ km\,s$^{-1}$).
On the other hand, the systemic velocity of the Orion Molecular Cloud (OMC) A is $\sim 8$ km\,s$^{-1}$ \citep{2019Natur.565..618P}, which is similar to the gases with the lower DCN/DCO$^+$ abundance ratios ($V_{\rm {LSR}} \approx 7.5-8.7$ km\,$^{-1}$).

In these different gas components, different deuterium fractionation pathways most likely proceed.
Based on comparisons of the velocity components, the gas components with the higher velocity component ($V_{\rm {LSR}} \approx 9.2-11.6$ km\,s$^{-1}$) are affected by the Trapezium cluster more significantly, and fluxes of high energy particles ($i.e.,$ UV photons and cosmic rays) should be higher compared to the behind dense OMC-1 with the lower velocity component ($V_{\rm {LSR}} \approx 7.5-8.7$ km\,s$^{-1}$).
The higher DCN/DCO$^+$ abundance ratios come from the gas affected by the expanding shell, while the lower abundance ratios are associated with the dense molecular clouds.
We will discuss detailed deuterium fractionation of these two species using our latest chemical simulation in Section \ref{sec:dis2}.

\subsection{Comparison with the chemical modeling} \label{sec:dis2}

\subsubsection{Chemical parameters used in \texttt{DNAUTILUS}}

We compare the observed abundances of DCN and DCO$^{+}$ as well as the DCN/DCO$^{+}$ ratios around the Orion KL region with modeled values.
To simulate the observed values around Orion KL, we have utilized our three-phase gas-grain chemical code called \texttt{DNAUTILUS.2.0}. 
This code calculates the time-dependent abundances of chemical species, including deuterated ones, by solving rate equations. 
It is an updated version of the \texttt{DNAUTILUS.1.0} astrochemical code introduced in \citet{2017MNRAS.466.4470M}.
The \texttt{DNAUTILUS.2.0} is capable of studying deuterium fractionation in both two- and three-phase modes, with or without ortho and para spin-states of major hydrogenated species, such as H$_2$, H$_2^+$, and H$_3^+$, and their isotopologues. 
In this work, we employed the three-phase version of \texttt{DNAUTILUS 2.0} without spin-state chemistry for computational efficiency and simplicity when running the parameter space summarized in Table \ref{tab:para1}.
This version of \texttt{DNAUTILUS 2.0} encompasses all the physicochemical processes and their corresponding equations derived from the state-of-the-art publicly available astrochemical code \texttt{NAUTILUS}, as detailed in \citet{2016MNRAS.459.3756R}. 

\texttt{DNAUTILUS 2.0} includes 1606 gas species, 1472 grain-surface species, and 1472 grain-mantle species, connected by 83,715 gas-phase reactions, 10,967 reactions on grain surfaces, and 9,431 reactions in the grain mantles. 
This version was generated by applying our well-tested deuteration routine to the gas, grain, and mantle reaction networks of \texttt{NAUTILUS}, where deuterons substitute for protons in the reactions, and branching ratios are calculated assuming complete scrambling. 
This methodology is similar to the one described in \citet{2013ApJS..207...27A}.

The binding energies for neutral deuterated species are assumed to be similar to their non-deuterated counterparts, following previous studies \citep{2013AA...550A.127T,2013ApJS..207...27A,2017MNRAS.466.4470M}. 
This is unavoidable due to the limited data available on the binding energies of deuterated molecules on grain surfaces. 
The binding energy of DCN is set at 2050 K, similar to the standard value of HCN as considered in \texttt{NAUTILUS} \citep{2016MNRAS.459.3756R}. 

For initial abundances, the species are assumed to be initially in atomic form, similar to diffuse clouds, except for hydrogen and deuterium, which are initially in H$_2$ and HD forms, respectively. 
The initial elemental abundances are summarized in Table \ref{tab:para1}. 

\begin{deluxetable}{ll}
\tablecaption{Physical parameters and initial abundances \label{tab:para1}}
\tablewidth{0pt}
\tablehead{
\colhead{Parameter} & \colhead{Value} 
}
\startdata
$T$ [K] & 10, 20, 30, 40, 50 \\
n$_{\rm {H}}$ [cm$^{-3}$] & $10^3-10^5$ \\
$\zeta$ [s$^{-1}$] & $10^{-17} - 10^{-15}$ \\
\hline
\multicolumn{2}{c}{Initial elemental abundances}\\
\multicolumn{2}{c}{Element \quad Abundance relative to H} \\
\hline
H$_2$ &   $0.5$ \\
He    &   $9.00\times10^{-2}$ \\
N     &   $6.20\times10^{-5}$ \\
O     &   $2.40\times10^{-4}$ \\
C$^+$ &   $1.70\times10^{-4}$ \\
S$^+$ &   $8.00\times10^{-8}$ \\
Si$^+$&   $8.00\times10^{-9}$ \\
Fe$^+$&   $3.00\times10^{-9}$ \\
Na$^+$&   $2.00\times10^{-9}$ \\
Mg$^+$&   $7.00\times10^{-9}$ \\
P$^+$ &   $2.00\times10^{-10}$ \\
Cl$^+$&   $1.00\times10^{-9}$ \\
F     &   $6.68\times10^{-9}$ \\
HD    &   $1.60\times10^{-5}$ \\
\enddata
\end{deluxetable}

\subsubsection{Physical parameters used in \texttt{DNAUTILUS}}

The physical model we adopted for the simulation is similar to that of \citet{2015AA...578A..55S}. We chose to explore a broader parameter space, including temperature ($T$), density (n$_{\rm {H}}$), and the cosmic-ray (CR) ionization rate ($\zeta$).
The physical parameter spaces used in this paper are listed in Table \ref{tab:para1}.
This choice is justified because isotopic ratios are observed in different regions of Orion KL with varying physical conditions.

The adapted density range covers the expanding shell ($\sim 10^3$ cm$^{-3}$) and OMC-1 ($10^4-10^5$ cm$^{-3}$) \citep{2019ApJ...881..130A,2019Natur.565..618P}.
The eastern side of the Orion KL hot core is primarily exposed to CRs and UV flux from OB massive stars in the Trapezium cluster, while the western side is heavily shielded.
Consequently, we can anticipate higher temperatures and secondary UV photon effects from CRs on the east side, which could impact the chemical outcomes.
We ran models with different CR ionization rates from $10^{-17}$ s$^{-1}$ to $10^{-15}$ s$^{-1}$ to simulate ionization fronts in the hot core generated by the OB massive stars in the Trapezium cluster and the shielded regions.

We maintained a fixed visual extinction of 300 mag, which approximately corresponds to the average (330 mag) and median (296 mag) values obtained from the analysis of five positions (Table \ref{tab:phy}) presented in this paper.
We kept the gas temperature and grain temperature constant.

\begin{deluxetable*}{cccc}
\tablecaption{Major formation and destruction reactions of DCN and DCO$^{+}$ at 0.2 Myr \label{tab:para3}}
\tablewidth{0pt}
\tablehead{
 \colhead{Densities (cm$^{-3}$)} & \colhead{$\zeta$ (s$^{-1}$)} & \colhead{DCN} & \colhead{DCO$^{+}$}
}
\startdata
Low (10$^{3}$ -- 10$^{4}$ cm$^{-3}$) & 2$\times$10$^{-17}$ &HDCN$^{+}$/DCNH$^{+}$ + e$^{-}$  $\rightarrow$ DCN + H & HCO$^{+}$ + D $\rightarrow$ DCO$^{+}$ + H \\
   & & DCN + H$_{3}$ $^{+}$ $\rightarrow$ HDCN$^{+}$/H$_{2}$CN$^{+}$ + H$_{2}$/HD & DCO$^{+}$ + e$^{-}$$\rightarrow$ CO + D \\
   & & & \\
   & 4$\times$10$^{-17}$ &HDCN$^{+}$/DCNH$^{+}$ + e$^{-}$ $\rightarrow$ DCN + H & HCO$^{+}$ + D $\rightarrow$ DCO$^{+}$ + H \\
   & & DCN + HCO$^{+}$ $\rightarrow$ DCNH$^{+}$ + CO & DCO$^{+}$ + e$^{-}$$\rightarrow$ CO + D \\
   & & & \\
\hline
High (10$^{4}$ -- 10$^{5}$ cm$^{-3}$) & 2$\times$10$^{-17}$ &HDCN$^{+}$/DCNH$^{+}$ + e$^{-}$  $\rightarrow$ DCN + H & HCO$^{+}$ + D $\rightarrow$ DCO$^{+}$ + H \\
   & & DCN + H$_{3}$ $^{+}$ $\rightarrow$ HDCN$^{+}$/H$_{2}$CN$^{+}$ + H$_{2}$/HD & DCO$^{+}$ + e$^{-}$$\rightarrow$ CO + D \\
   & & DCN  $\rightarrow$ DCN$\mathrm{_{ice}}$ &  \\
   & & & \\
   & 4$\times$10$^{-17}$ &HDCN$^{+}$/DCNH$^{+}$ + e$^{-}$ $\rightarrow$ DCN + H & HCO$^{+}$ + D $\rightarrow$ DCO$^{+}$ + H \\
   & & DCN + HCO$^{+}$ $\rightarrow$ DCNH$^{+}$ + CO & DCO$^{+}$ + e$^{-}$$\rightarrow$ CO + D \\
   & & DCN  $\rightarrow$ DCN$\mathrm{_{ice}}$ &  \\
   & & & \\
\enddata
\end{deluxetable*}

\subsubsection{Comparisons between the observations and models}

\begin{figure*}
 \begin{center}
  \includegraphics[bb = 0 30 900 500, scale = 0.35]{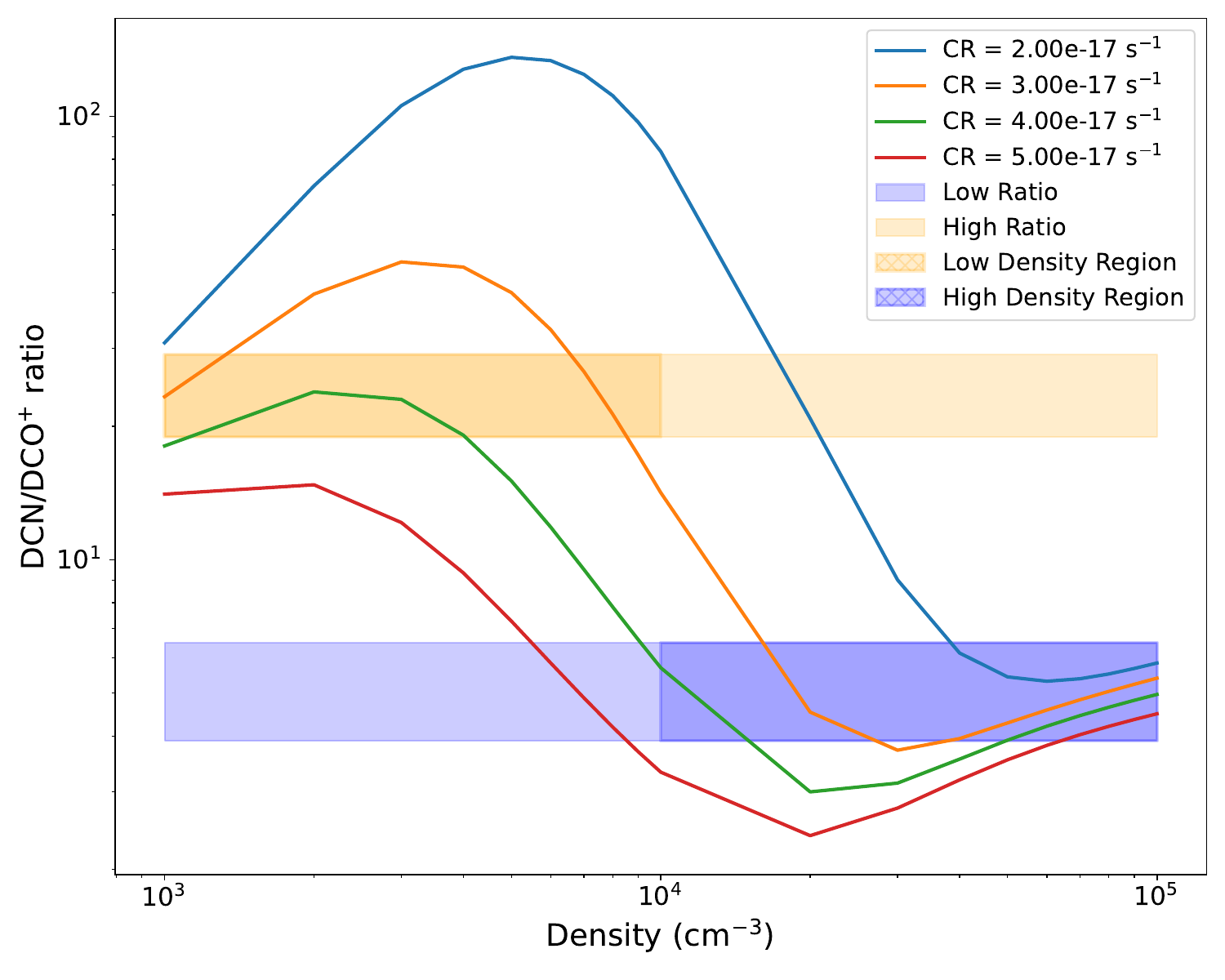}
 \end{center}
\caption{Comparisons between the modeled DCN/DCO$^{+}$ ratios at $T=40$ K with different CR ionization rates and the observed values. Shadow areas indicate the observed error ranges. The orange mesh area depicts the low-density region with a high DCN/DCO$^{+}$ ratio, while the blue mesh area depicts the high-density area with a low DCN/DCO$^{+}$ ratio.}
\label{fig:model}
\end{figure*}

We compared the modeled results with the temperatures of 10 -- 50 K to the observational results and confirmed that the models at 40 K reproduce the observational results best.
The temperature of 40 K agrees with the dust temperature in this region (see Table \ref{tab:phy}).
We thus show and discuss the modeled results at 40 K in the following parts.
Figure \ref{fig:model} shows the modeled DCN/DCO$^{+}$ abundance ratios along with the observed values at 0.2 Myr, which corresponds to the age of the expanding bubble constrained by \citet{2019Natur.565..618P}.
We found that the CR ionization rates larger than $5\times10^{-17}$ s$^{-1}$ cannot reproduce the observed abundance ratios and we exclude them in Figure \ref{fig:model}.
Major production and destruction reactions of DCN and DCO$^{+}$ for different densities and CR ionization rates at 40 K are summarized in Table \ref{tab:para3}.


Dissociative recombination reactions between molecular ions, DCNH$^{+}$ and HDCN$^{+}$, and electrons (e$^{-}$) produce DCN in low-density regions ($10^3-10^4$ cm$^{-3}$):

\begin{equation} \label{eq:recombination1}
{\rm {DCNH}}^+/{\rm {HDCN}}^+ + {\rm {e}}^- \rightarrow {\rm {DCN}} + {\rm {H}}.
\end{equation}

In such conditions, DCO$^{+}$ is efficiently destroyed through dissociative recombination with e$^{-}$:
\begin{equation} \label{eq:recombination2}
{\rm {DCO}}^+ + {\rm {e}}^- \rightarrow {\rm {CO}} + {\rm {D}}.
\end{equation}
Consequently, the DCN/DCO$^+$ abundance ratio increases in the low-density regions where the electron density is high.
In the dense regions ($\gtrsim 10^{4}$ cm$^{-3}$), freezing of DCN onto dust grains becomes important and decreases the gas-phase DCN abundances.
This decreases the DCN/DCO$^{+}$ ratios in high density regions.
In all these environments, the production of DCO$^{+}$ is primarily driven by the deuterium exchange reaction involving HCO$^{+}$ (reaction (\ref{eq:DCO2}); HCO$^+$ + D $\rightarrow$ DCO$^+$ + H). 


We found that density is the key parameter to produce the observed differences in the DCN/DCO$^+$ abundance ratio.
The CR ionization rate has little effect on the ratio in our assumed conditions.
The low observed DCN/DCO$^+$ ratios ($\sim4-7$) favor models with high densities of $10^4-10^5$ cm$^{-3}$, which are consistent with the density in OMC-1 \citep{2019Natur.565..618P}.
This situation is consistent with observational results, where gas components with $V_{\rm {LSR}}\approx8$ km\,s$^{-1}$, unaffected by the expanding shell, exhibit the low DCN/DCO$^+$ ratios.
On the other hand, the high observed abundance ratios (DCN/DCO$^+ \approx 22-30$) are well replicated by models with low densities of $10^3-10^4$ cm$^{-3}$. 
Such lower densities agree with conditions of the expanding shell \citep{2019ApJ...881..130A}.
This is consistent with the observational results that the high abundance ratios are found in  $V_{\rm {LSR}}\approx9.2-11.6$ km\,s$^{-1}$, likely influenced by the expanding shell generated by the most massive star in the Trapezium cluster.

\section{Conclusions} \label{sec:con}

We have obtained the $1.5\arcmin$ (0.18 pc)-square maps of the DCN and DCO$^+$ lines toward the Orion KL region with the B4R system installed on the Large Millimeter Telescope.
The main conclusions of this paper are as follows.

\begin{enumerate}
\item The DCN emission shows a peak at the hot core position, whereas that of DCO$^+$ does not emit there. 
The DCO$^+$ emission is enhanced in the west side from the hot core which is well shielded from the UV photons and CRs from the OB stars in the Trapezium cluster.

\item We have derived the DCN/DCO$^+$ abundance ratios at the three representative positions where both species have been detected. 
The ratios are divided into two groups; DCN/DCO$^+$$\approx 4-7$ and $\approx22-30$. 
The former is associated with the gas components of $V_{\rm {LSR}}\approx7.5-8.7$ km\,s$^{-1}$, which is consistent with the systemic velocity of OMC-1.
On the other hand, the gas components with the higher abundance ratios have velocity components of 9.2--11.6 km\,s$^{-1}$, which are likely affected by the expanding shell produced by the most massive star in the Trapezium cluster.

\item The two distinct abundance ratios can be reproduced by our models. 
Models featuring high-density 
provide a better explanation for the low observed DCN/DCO$^+$ ratio. 
On the other hand, models with low-density 
reproduce the high observed abundance ratio. 
This condition is consistent with the presence of an expanding shell caused by the OB stars. 
\end{enumerate}


\begin{acknowledgments}
This paper makes use of data taken by the Large Millimeter Telescope Alfonso Serrano (LMT) in Mexico.
The LMT project is a joint effort of the Instituto Nacional de Astr\'{o}fisica, \'{O}ptica, y Electr\'{o}nica (INAOE) and the University of Massachusetts at Amherst (UMASS).
We also appreciate the support of the technical staff and the support scientists of the LMT during the commissioning campaign of the B4R.
Data analysis was in part carried out on the Multi-wavelength Data Analysis System operated by the Astronomy Data Center (ADC), National Astronomical Observatory of Japan.
K.T. is supported by JSPS KAKENHI grant No. JP20K14523 and 21H01142. 
T.Y. is supported by JSPS KAKENHI grant No. JP22J22889.
We thank the anonymous referee whose comments helped improve the paper.
\end{acknowledgments}

\vspace{5mm}
\facilities{LMT, B4R}

\software{Common Astronomy Software Applications package \citep[CASA;][]{2022PASP..134k4501C}, CASSIS \citep{2015sf2a.conf..313V}}


\end{document}